\documentclass[prb,aps,twocolumn,showpacs,epsf]{revtex4}
\begin{document}

\title{Understanding cuprate superconductors with spontaneous nodal gap generation}
\author{Guo-Zhu Liu$^{1}$ and Geng Cheng$^{2,3}$ \\
$^{1}${\small {\it Department of Mathematics, University of Science and Technology of China,
Hefei, Anhui, 230026, P.R. China }}\\
$^{2}${\small {\it CCAST (World Laboratory), P.O. Box 8730,
Beijing 100080,
P.R. China }}\\
{\small {\it $^{3}$Department of Modern Physics, University of
Science and Technology of China, Hefei, Anhui, 230026, P.R.
China}}}

\begin{abstract}
We study the spontaneous gap generation for gapless nodal fermions
within an effective gauge field theory of high temperature
superconductors. When superconductivity appears, the gauge boson
acquires a finite mass via Anderson-Higgs mechanism. Spontaneous
nodal gap generation takes place if the gauge boson mass $\xi$ is
zero or less than a critical value $\xi_{c}$ but is suppressed by
a larger gauge boson mass. The generated nodal gap prohibits the
appearance of low-energy fermion excitations and leads to
antiferromagnetic order. Using the fact that gauge boson mass
$\xi$ is proportional to superfluid density and doping
concentration, we build one mechanism that provides a unified
understanding of the finite single particle gap along the nodal
direction in lightly doped cuprates, the competition and
coexistence of antiferromagnetism and superconductivity, and the
thermal metal-to-insulator transition from the superconducting
state to the field-induced normal state in underdoped cuprates.
\end{abstract}

\maketitle


\section{introduction}

Recently, angle-resolved photoemission spectroscopy (ARPES)
measurements were performed on lightly doped cuprates and found a
full gap over the whole Brillouin zone \cite{kmShen}. This finding
is unexpected since the $d$-wave symmetry of the gap/psedogap of
high temperature superconductors has widely accepted \cite{Tsuei}
and previous extensive ARPES measurements always revealed gapless
excitations in the $\left(\pm \pi/2,\pm \pi/2 \right)$ direction
\cite{zxShen}. The nodal gap decreases upon doping and disappears
when superconductivity emerges as the ground state. In some
materials with superconductivity being its ground state, the nodal
gap is also observed. This indicates that there seems to be a
competition and possible coexistence between the nodal gap and
superconductivity. This behavior is quite similar to the evolution
of orders with doping concentration. Antiferromagnetism is the
ground state of undoped and lightly doped cuprate superconductors,
but disappears when superconductivity emerges, thus there is a
competition between antiferromagnetism and superconductivity.
Because of this competition, when superconductivity is suppressed,
say by magnetic fields, antiferromagnetism has a chance to appear
in a superconductor. The field-induced local antiferromagnetic
order has been confirmed by recent neutron scattering and scanning
tunnel microscopic (STM) experiments \cite{Lake, Mitrovic,
Hoffman, Khaykovich}. In particular, antiferromagnetic order is
found not only in the Abrikosov vortices but also in the
superconducting region around the vortices. This finding strongly
supports the coexistence of antiferromagnetism and
superconductivity. On the other hand, extensive heat transport
measurements show that a residual linear term exists in the whole
superconducting region at $T \rightarrow 0$, but it decreases with
decreasing doping concentration and approaches zero as
superconductivity disappears \cite{Taillefer, Chiao, Hawthorn,
Takeya, Sutherland}. When an external magnetic field is present
perpendicular to the CuO$_{2}$ plane, the thermal conductivity of
underdoped cuprates decreases with increasing magnetic field and
finally vanishes when the magnetic field is beyond the up critical
field $H_{c2}$. Thus cuprate superconductors exhibit a thermal
metal-to-insulator transition upon going from the superconducting
state to the field-induced normal state \cite{Hawthorn}.

We believe that the above phenomena are universal to all high
temperature supercodncutors and more importantly they are governed
by the $same$ physical mechanism. In this paper, we argue that all
these experimental results can be understood if the gapless nodal
fermions acquire a finite gap. Such a spontaneous gap generation
for nodal fermions is achieved by coupling the fermions to a gauge
field which naturally appears as a result of strong electron
correlation effect when we go beyond the slave-boson mean-field
treatment of $t$-$J$ model.

Spin-charge separation and emergent gauge fluctuation are two key
concepts in our scenario. When spin and charge degrees of freedom
are separated, the excitations are spin-carrying spinons and
charge-carrying holons rather than ordinary electrons. The pairing
of spinons is responsible for the observed $d$-wave energy
gap/pseudogap. The $d$-wave gap vanishes at the nodes, so the
low-energy fermion excitations are effectively gapless and hence
can be described by relativistic massless Dirac fermions.
Superconductivity is realized once the holons undergo Bose
condensation at low temperatures. The spinons are connected to the
holons due to the exchange of an emergent gauge field although
they do not interact directly. Thus, the low-energy behavior is
dominated by an interacting system consisting of gapless Dirac
fermions, holons and an emergent gauge field.

A finite nodal gap is generated spontaneously when the gauge field
binds gapless nodal fermions into stable fermion-anti-fermion
pairs. Exciting single fermions from the nodal direction of the
Brillouin zone requires a finite energy cost which is responsible
for the finite nodal gap observed in ARPES experiments. Since the
single particle spectrum is fully gapped in the whole momentum
region, no free fermions can exist at low temperatures and
consequently there should not be a linear term for the thermal
conductivity, which is present in a pure $d$-wave superconductor
with gap nodes. Spontaneous nodal gap generation can take place if
the holons are absent or free. When superconductivity emerges as
the ground state, the local gauge symmetry is broken by the holon
condensation and the gauge boson becomes massive via
Anderson-Higgs mechanism. We found a critical value for the gauge
boson mass. Spontaneous nodal gap generation takes place if the
gauge boson mass is less than the critical value but is suppressed
when the gauge boson mass becomes larger than the critical value.
Then there is a competition between spontaneous nodal gap
generation and superconductivity. The linear term for thermal
conductivity in the superconducting state observed by heat
transport measurements reflects the suppression of spontaneous
nodal gap generation by superconductivity. Another important
consequence of spontaneous nodal gap generation is that it
enhances the antiferromagnetic spin correlation greatly and
actually leads to long-range order. Thus the antiferromagnetic
order should have the same doping dependence with the nodal
fermion gap: it is present in low doping region and disappears in
high doping region after superconductivity emerges. This is
exactly what happens in cuprates superconductors. Since the gauge
boson mass is proportional to the superfluid density, the
coexistence of spontaneous nodal gap generation and a small gauge
boson mass leads to an very important result that the
antiferromagnetic order can coexist with superconductivity when
the superfluid density is less than the critical value. Thus the
spontaneous nodal gap generation provides a unified explanation of
the experimental results mentioned at the beginning of this paper.

This paper is organized as follows: In Sec. (II), we first give a
brief review on spin-charge separation, slavo-boson treatment of
$t$-$J$ model and the gauge theory approach to cuprate
superconductors, then we discuss chiral symmetry breaking in the
presence of massive gauge boson and estimate the critical gauge
boson mass which separates chiral symmetric and symmetry broken
phases. In Sec. (III), we discuss the explanation of the ARPES
experiments, the heat transport behavior and the relationship
between antiferromagnetism and superconductivity respectively. The
paper ends with a summary.

\section{dynamical gap generation for gapless nodal fermions}

Shortly after the discovery of high temperature superconductors,
it was correctly recognized that these materials are doped Mott
insulators and hence can not be properly understood without
considering the strong correlation effect. Anderson
\cite{Anderson87} proposed that due to quantum fluctuations the
ground state of undoped cuprates is more probably some kind of
quantum liquid of spin singlets, called resonating valence bond
(RVB) state. The investigation of RVB state was carried out within
the $t-J$ model which is derived from the more general three-band
Hubbard model \cite{Rice}. The strong correlation nature of
cuprate superconductors is reflected in a no-double occupancy
constraint which says that there is no more than one electron at
one lattice site due to the strong Coulomb repulsion between
electrons. By decomposing the electron operator
$c_{i\sigma}^{\dag}$ to the product of a spinon operator
$f_{i\sigma}^{\dag}$ (neutral fermion) and a holon operator
$b_{i}$ (charged boson)
\begin{equation}
c_{i\sigma}^{\dag}=f_{i\sigma}^{\dag} b_{i},
\end{equation}
the no-double occupancy constraint can be written as
\begin{equation}
\sum_{\sigma} f_{i\sigma}^{\dag} f_{i\sigma}+ b_{i}^{\dag}b_{i} =
1
\end{equation}
which is easier to be treated analytically. This decomposition is
the crucial step underlying the so-called slave-boson mean field
treatment of $t-J$ model. A four-fermion interaction term appears
in the $t$-$J$ model after replacing electron operators with
spinon operators and holon operators. This term can be treated by
introducing three order parameters $\chi_{ij}=\left<
f_{i\sigma}^{\dag} f_{i\sigma}\right>$, $\Delta_{ij}=\left<
f_{i\uparrow}f_{j\downarrow} -
f_{i\downarrow}f_{j\uparrow}\right>$, and
$\eta_{ij}=b_{i}^{\dag}b_{j}$. A phase diagram can be obtained
based on this mean field approach  \cite{Baskaran, Kotliar}. The
flux phase was found to be locally stable \cite{Affleck} and very
interesting for its applicability to cuprate superconductors. Due
to its $d$-wave spinon gap symmetry, the low-energy excitations
are actually gapless Dirac fermions \cite{Affleck, Ioffe}. The
quantum fluctuations around this mean field state includes a
massless U(1) gauge field. The low-energy effective behavior of
cuprate superconductors thus can be well described by the
three-dimensional quantum electrodynamics (QED$_{3}$).

Although the U(1) formulation captures some important properties
of high temperature superconductors, it was shown by Wen and Lee
\cite{Wen96} not to connect smoothly to the hall-filling material
in which an exact SU(2) local symmetry was found. An SU(2)
treatment of the 2D $t$-$J$ model was constructed to describe the
physics of undoped and underdoped cuprates in a unified way
\cite{Wen96}. This treatment gives rise to several mean-field
phase diagrams among which the staggered flux phase is applicable
to cuprate superconductors. Taking quantum fluctuations into
account leads to a low-energy effective theory \cite{Kim99} that
consists of a massless U(1) gauge field, a two-component bosons,
and a massless fermions excited from the gap nodes of $d$-wave
spinon pairs.

\subsection{Dynamical chiral symmetry breaking}

If the fermions are massless the theory respects a chiral
symmetry. However, the QED$_{3}$ theory has a rather peculiar
property that the massless fermions can acquire a dynamically
generated mass when its flavor is less than a critical value. The
mass term of fermions spontaneously breaks the chiral symmetry
which leads to a massless Goldstone boson according to the
Goldstone theorem.

At half-filling, the low energy physics is dominated by the
interaction of nodal fermions with U(1) gauge fields \cite{Kim99}
\begin{equation}
{\cal L}_{F}=\sum_{\sigma=1}^{N}\overline{\psi}_{\sigma}
v_{\sigma, \mu} \left( \partial_{\mu }-ia_{\mu} \right)
\gamma_{\mu}\psi_{\sigma}.
\end{equation}
The Fermi field $\psi_{\sigma}$ is a $4\times 1$ spinor
representing the gapless nodal fermions. The $4 \times 4$ $\gamma
_{\mu}$ matrices obey the algebra, $\lbrace
\gamma_{\mu},\gamma_{\nu} \rbrace=2\delta_{\mu \nu}$, and for
simplicity, we let $v_{\sigma, \mu}=1$ ($\mu,\nu=0,1,2$). At
half-filling case, there are no holons or the holons are all
confined because of the presence of a very large charge gap.

Chiral symmetry breaking is a nonperturbative phenomenon and can
not be obtained within any finite order of the perturbation
expansion. The standard approach to this problem is to solve the
self-consistent Dyson-Schwinger (DS) equation for the fermion
self-energy. The inverse fermion propagator is written as
$S^{-1}(p)=i\gamma \cdot p A \left( p^{2} \right)+\Sigma \left(
p^{2} \right)$, $A(p^{2})$ is the wave-function renormalization
and $\Sigma(p^{2})$ the fermion self-energy. The propagator of a
massless fermion is simply $S^{-1}(p)=i\gamma \cdot p$. $A(p^{2})$
and $\Sigma (p^{2})$ appear due to the renormalization effect
caused by interaction with gauge field. The self-energy function
$\Sigma (p^{2})$ represents the interaction induced fermion mass
and is determined by a set of DS integral equations. If the DS
equation for $\Sigma(p^{2})$ has only vanishing solutions, the
fermions remain gapless and the Lagrangian respects the chiral
symmetries $\psi \rightarrow \exp (i\theta \gamma _{3,5})\psi$,
with $\gamma _{3}$ and $\gamma _{5}$ two $4 \times 4$ matrices
that anticommute with $\gamma _{\mu}$ ($\mu=0,1,2$). If the DS
equation for $\Sigma(p^{2})$ develops a squarely integrable
nontrivial solution \cite{Liu2001}, then the originally massless
fermions acquire a finite mass. The DS equation is extremely
complicated and hence can never be treated without making proper
approximations. To the lowest order \cite{Appelquist88}, we take
$A=1$ neglecting all higher-order corrections and approximate the
vertex function by the bare $\gamma_{\mu}$. After making these
approximations, the DS equation has the form
\begin{equation}
\Sigma(p^{2})=\int\frac{d^{3}k}{(2\pi)^{3}}\frac{\gamma^{\mu}D_{\mu\nu}(p-k)\Sigma(k^{2})\gamma^{\nu}}{k^{2}
+\Sigma^{2}(k^{2})}.
\end{equation}
Appelquist $et$ $al.$ showed \cite{Appelquist88, note} that this
DS equation has solutions only for $N < N_{c}=32/\pi^{2}$. Since
the physical fermion flavor is $2$ representing the number of spin
component, the massless fermions actually acquire a finite mass.
This mechanism is called dynamical chiral symmetry breaking and
has been studied for many years in particle physics as a possible
mechanism to generate fermion mass without introducing annoying
Higgs bosons. The presence of a critical fermion flavor can be
understood as follows. The effective coupling of gauge field is
proportional to $1/N$. It is strong enough to form fermion pairs
when the physical fermion flavor is less than $N_{c}$, while for
large $N$ the coupling is too weak.

\subsection{Effect of gauge boson mass on chiral structure}

When holes are doped into the CuO plane, the holons are excited
and hence an additional coupling between holons and the gauge
field appears. We have shown \cite{LiuWF} in a gauge invariant way
that the gapless fermions can acquire a finite gap when the
additonal holons are not Bose condensed. When superconductivity
emerges, the gauge boson becomes massive. In order to understand
the properties of nodal fermions we need to investigate the effect
of the gauge boson mass on chiral symmetry breaking.

We write the action of bosons in the standard Ginsberg-Landau form
\begin{equation}
{\cal
L}_{B}=\frac{1}{4m_{b}}\left|\left(\partial_{\mu}-ia_{\mu}\right)b\right|^{2}-\alpha
\left| b \right| ^{2}-\frac{\beta }{2}\left| b \right| ^{4}.
\end{equation}
Note that this is not the popular model that has been extensively
studied in the literature \cite{Kim99}. In previous treatment, a
non-relativistic model has been used to describe the interaction
of holons and gauge fields. In such a model, the density
fluctuations of holons screen the temporal component of the gauge
field which becomes massive and hence is ignored. However, this
treatment destroys the gauge invariance of the theory: the result
obtained in the Landau gauge is qualitatively different from that
in the Feynman gauge \cite{LiuWF}. This inconsistency might be a
result of using an inappropriate Lagrangian for the holons. At
present, it is not possible to derive an effective action for the
holons rigorously. In this paper, we use the relativistic scalar
QED to describe the holons because it is the most general field
theoretic model for a scalar field and it can lead to a gauge
invariant critical fermion flavor.

In the Lagrangian, $\beta$ is always positive while $\alpha$ can
be positive or negative corresponding to normal and
superconducting states, respectively. For $\alpha
>0$, the holons are free and the Lagrangian $L_{B}$ is invariant under
the local U(1) gauge transformation
\begin{equation}
a_{\mu }\longrightarrow a_{\mu }-\partial _{\mu }\theta
\end{equation}
\begin{equation}
b \longrightarrow e^{i\theta }b,
\end{equation}
where $\theta \left( x\right)$ is an arbitrary function. When the
holons undergo Bose condensation, $\alpha$ becomes negative and
the ground state occurs at
\begin{equation}
\left\langle b \right\rangle = b_{0}=\sqrt{-\frac{\alpha}{\beta}}.
\end{equation}
Thus, the local gauge symmetry is spontaneous broken due to ground
state degeneracy. We could write the holon field as
\begin{equation}
b \left( x\right) = b_{0}e^{i\theta \left( x\right)},
\end{equation}
where $\theta$ is the phase of the order parameter which is just
the gapless Goldstone mode associated with the spontaneous gauge
symmetry breaking. The appearance of this gapless mode used to a
big puzzle to physicists since no such modes had been observed in
superconductors. This inconsistency could be eliminated by the
Anderson-Higgs mechanism \cite{LiuAH}. Its essential idea is to
perform the following gauge transformation
\begin{equation}
a_{\mu }\longrightarrow a_{\mu}+\partial _{\mu }\theta,
\end{equation}
taking advantage of gauge freedom of the theory. Then the
Goldstone mode $\theta$ is removed and there appears a term as
follows
\begin{equation}
\frac{1}{m_{b}}b _{0}^{2}a_{\mu}a^{\mu}.
\end{equation}
It is easy to see that the gauge boson now acquires a finite mass
\begin{equation}
\xi=\frac{1}{m_{b}}\rho_{s}
\end{equation}
after absorbing the gapless Goldstone mode. This is the famous
Anderson-Higgs mechanism which when generalized to non-Abelian
gauge theories constitutes the foundation of the electro-weak
Standard Model. The physical meaning of the finite gauge boson
mass can be seen from the equation for magnetic field
\begin{equation} \nabla ^{2}{\bf B}=\xi {\bf B}.
\end{equation}
Comparing this with the standard London equation, we know that
$\xi = \lambda_{L}^{-2}$ with $\lambda_{L}$ the London penetration
depth. The solution of this equation is an exponentially damping
function indicating that the magnetic field can penetrate the
superconductors only to a finite depth equal to the inverse gauge
boson mass.

After the gauge boson acquires a finite mass, its coupling
strength is weakened and might not be able to form fermion pairs.
To testify the correctness of this naive expectation, we now
investigate the effect of a finite gauge boson mass on chiral
symmetry breaking by studying the DS equation with massive gauge
boson propagator and seek the critical coupling constant. The
gauge boson propagator in Landau gauge is
\begin{equation}
D_{\mu\nu}(q)=D_{T}(q^{2})\left(\delta_{\mu\nu}-\frac{q_{\mu}q_{\nu}}{q^{2}}\right),
\end{equation}
with
\begin{equation}
D_{T}^{-1}(q^{2})=q^{2}\left[1+\pi(q^{2})\right]+\xi^{2}.
\end{equation}
Let $q$ be the gauge boson momentum, then
$q^{2}=(p-k)^{2}=p^{2}+k^{2}-2pk\cos{\theta}$. $\pi(q^{2})$ is the
vacuum polarization of the gauge boson, which is originally
introduced to overcome the infrared divergence. In the present
case, the gauge field has no kinetic energy term and its dynamics
comes from integrating out matter fields. As a result, only
$\pi(q^{2})$ appear in $D_{T}(q^{2})$, so
\begin{equation}
D_{T}^{-1}(q^{2})=q^{2}\pi(q^{2})+\xi^{2}.
\end{equation}
The vacuum polarization consists of two parts corresponding to
fermion contribution $\pi_{F}$and holon contribution $\pi_{B}$
respectively, which are
\begin{equation}
\pi_{F}(q^{2})=\frac{N}{8\left|q\right|},
\end{equation}
\begin{equation}
\pi_{B}(q^{2})=\frac{1}{8\left|q\right|}
\end{equation}
to the one-loop approximation (Note that the holon mass is ignored
in the polarization $\pi_{B}$ since it does not affect the chiral
structure \cite{LiuWF}). Adding them up leads to the total vacuum
polarization
\begin{equation}
\pi(q^{2})=\frac{N+1}{8\left|q\right|}.
\end{equation}
Then we have
\begin{eqnarray}
D_{T}^{-1}(q^{2})=q^{2}\pi(q^{2})+\xi^{2} = \frac{N+1}{8}(q+\eta),
\end{eqnarray}
with
\begin{equation}
\eta=\frac{8\xi^{2}}{N+1}.
\end{equation}
Now we can write down the propagator of gauge boson explicitly
\begin{equation}
D_{\mu\nu}(q)=\frac{8}{(N+1)(\left|q\right|+\eta)}\left(
\delta_{\mu\nu}-\frac{q_{\mu}q_{\nu}}{q^{2}}\right).
\end{equation}
We now substitute this expression into Eq. (4). After performing
angular integration and introducing an ultraviolet cutoff
$\Lambda$ we have
\begin{eqnarray}
\Sigma(p^{2})&=&\lambda\int^{\Lambda}_{0} dk\frac{k\Sigma(k^{2})}{k^{2}+\Sigma^{2}(k^{2})} \nonumber \\
&&\times \frac{1}{p}\left(p+k-\left|p-k\right|-\eta\ln\left(\frac{p+k+\eta}{\left|p-k\right|+\eta}\right)\right),
\end{eqnarray}
where $\lambda=4/(N+1)\pi^{2}$ serves as an effective coupling
constant \cite{Liu2003}.

This integral equation can be investigated by bifurcation theory
and parameter imbedding method. In order to obtain the bifurcation
points we need only to find the eigenvalues of the associated
Fr\^{e}chet derivative of the nonlinear DS equation
\cite{Liu2003}. Those eigenvalues that have odd multiplicity are
the bifurcation points. The first bifurcation point is just the
critical coupling strength at which a nontrivial solution of the
DS equation develops \cite{Liu2003}. Once we obtained the critical
coupling constant, then we can get the critical fermion flavor
that separates the chiral symmetry breaking phase and the chiral
symmetric phase.

Numerical calculations \cite{Liu2003} found that the critical
fermion number is a monotonically increasing function of
$\Lambda/\eta$. It conforms the naive expectation that a finite
mass of the gauge boson is repulsive to gap generation for
fermions. For small $\Lambda/\eta$ the critical number is smaller
than physical fermion number 2, so fermions remain gapless. When
$\Lambda/\eta$ increases, the critical number increases
accordingly and finally approaches a constant value larger than
$2$. Thus we can conclude that the spontaneous gap generation for
fermions takes place when the gauge boson mass is zero and very
small but is destroyed when the gauge boson mass is larger than a
critical value. Including the wave function renormalization
$A(p^{2})$ does not change this behavior, but changes the critical
value quantitatively. We have performed calculations in an
appropriate non-local gauge after taking $A(p^{2})$ into account
and found that $\frac{\eta_{c}}{\Lambda}$ is between $10^{-4} eV$
and $10^{-3} eV$. Although it is hard to determine its exact
value, we will show that this uncertainty only has minor influence
on the critical values of observable quantities such as the
superfluid density and the doping concentration.

We know from these results that there is a competition between two
kinds of spontaneous symmetry breaking: chiral symmetry breaking
and spontaneous gauge symmetry breaking. Chiral symmetry breaking
is associated with the mass generation of massless fermions, while
spontaneous gauge symmetry breaking is caused by holon
condensation and generates a mass for the gauge boson. If the
gauge boson mass is less than the critical value but nonzero,
there is a coexistence of chiral symmetry breaking and spontaneous
gauge symmetry breaking.

\subsection{Critical point between chiral symmetric and symmetry breaking phases}

We now would like to discuss the critical value for the gauge
boson mass since it plays a crucial role in determining the
transition between chiral symmetric and symmetry breaking phases.
The gauge boson mass is not a physical quantity that can be
observed by experiments. In order to explain experiments with our
mechanism, we should make a connection between the gauge boson
mass with the superfluid density and the doping concentration,
which are important physical quantities in describing the
superconductors.

The relationship between gauge boson mass with superfluid density
we obtained in Eq. (12) is very useful in describing the effect of
superconducting condensation on spontaneous gap generation for the
nodal quasiparticles. Now we wish to relate the gauge boson mass
to doping concentration $\delta$. Careful London penetration depth
measurements and optical conductivity experiments \cite{Orenstein}
showed that the superfluid density is proportional to the doping
concentration, that is
\begin{equation}
\rho_{s}\left( T=0 \right)=\frac{\delta}{a^{2}},
\end{equation}
where $a$ is the lattice constant. The optical conductivity
experiments reflects the response of the system to electromagnetic
field, instead of the internal gauge field. This seems to be an
obstacle to use the above equation. However, we expect it works
well for both the internal gauge field and the electromagnetic
field since they have the same gauge structure. Thus, we obtain
the relationship between gauge boson mass and doping concentration
\begin{equation}
\delta=m_{b}a^{2}\xi.
\end{equation}
Based on this formula, the doping dependence of many physical
quantities can be described by their dependence of the gauge boson
mass. Note that this property is special to cuprate
superconductors which are believed to be doped Mott insulators.

Next we would like to calculate the critical value of the doping
concentration which separates the chiral symmetric and symmetry
breaking phases. As will be seen in the following discussions,
this is an important quantity in understanding experiments. Our
calculation of DS equation has given the critical gauge boson
mass. From the numerical results in the nonlocal gauge
\cite{Liu2003}, we know that $\frac{\eta_{c}}{\Lambda}$ is between
$10^{-4} eV$ and $10^{-3} eV$. The continuum U(1) gauge theory is
the effective low-energy theory of high temperature
superconductors, hence the lattice provides a natural ultraviolet
cutoff. However, we can also choose $\alpha=(N+1)/8$ as the
ultraviolet cutoff since in QED$_{3}$ all physical quantities
damps rapidly above this energy scale \cite{Appelquist88}.
Remember that we have defined $\eta=8\xi^{2}/(N+1)$. Putting all
these together, we get the critical doping concentration
\begin{equation}
\delta_{c}=\frac{\xi_{c}}{2\left(2m_{b}a^{2}\right)^{-1}}.
\end{equation}
The mass of the holons is determined by the hopping integral $t$
in the $t$-$J$ model. In the tight-binding treatment, we have
\begin{equation}
\left(2m_{b}a^{2}\right)^{-1}=t_{h}=0.122 {\text eV},
\end{equation}
which was obtained by Lee and Wen \cite{Lee97} in the case of
YBCO$_{6.95}$. If $\frac{\eta_{c}}{\Lambda}=10^{-4} eV$ then the
critical doping concentration is $\delta_{c}=0.03$; while if
$\frac{\eta_{c}}{\Lambda}=10^{-3} eV$, then $\delta_{c}=0.05$.
Since the antiferromagnetic order disappears generally at $0.03$,
the critical $\delta_{c}$ we obtained is in good agreement with
experiments and the inability in determining the exact value of
$\frac{\eta_{c}}{\Lambda}$ does not affect the reliability of our
conclusion.

Spontaneous gap generation takes place for doping concentration
less than $\delta_{c}$, no matter the ground state is
superconducting or not. For doping concentration larger than
$\delta_{c}$, if the ground state is not superconductivity, then
spontaneous gap generation also takes place; however, if the
ground state is superconducting, then spontaneous nodal gap
generation is suppressed by superconductivity. In the
superconducting state with doping concentration higher than
$\delta_{c}$, there is possibility that the superfluid density is
reduced to below its critical value $\rho_{sc}$ by some means
other than decreasing doping. If this really happens, then
spontaneous gap generation can also take place. This coexistence
of superconductivity and spontaneous gap generation is very
important and will be discusses in the next section.

\section{Explaining experiments with spontaneous nodal gap generation}

Since we know the effect of holons and gauge boson mass on the gap
structure of nodal fermions, we are equipped for understanding the
exotic experimental findings we have mentioned in the
Introduction. These experiments include the finite single particle
gap along the nodal direction, the low temperature quasiparticle
heat transport, the doping dependence of antiferromagnetism and
its relationship with superconductivity. The common feature of
these phenomena is that they are all dominated by the behavior of
nodal quasiparticles. It is thus not surprised that they can be
understood by one single physical mechanism. Although our
mechanism is still rather qualitative, its efficiency in
explaining these experimental results gives us confidence that it
does capture some essential physics of hign temperature
superconductors.

\subsection{Finite energy gap in nodal direction}

Now we discuss the application of our theory to ARPES experiments.
In RVB theories, all the single particle gaps are caused by the
pair formation of spinons. The spinon gap exists at any doping
concentration that is less than optimal doping and has $d$-wave
symmetry. ARPES measurements play an significant role in studying
the gap symmetry since it is momentum dependent. Most previous
ARPES measurements have been limited to optimally doped
Bi$_{2}$Sr$_{2}$CaCu$_{2}$O$_{8+x}$ because these materials have
ideal surfaces which are required by ARPES measurements. However,
to understand high temperature superconductivity, it is necessary
to know the electronic structure of underdoped, lightly doped and
undoped cuprates.

Recently, ARPES measurements have been performed in the lightly
doped cuprates including hole-doped La$_{2-x}$Sr$_{x}$CuO$_{4}$
and Ca$_{2-x}$Na$_{x}$CuO$_{2}$Cl$_{2}$ and electron-doped
Nd$_{2-x}$Ce$_{x}$CuO$_{4}$ \cite{kmShen}. The spectra from
Ca$_{2-x}$Na$_{x}$CuO$_{2}$Cl$_{2}$ reveals that a finite gap
exists in the nodal direction at $x=0.05$ and becomes smaller with
increasing doping concentration. It closes at $x=0.12$
corresponding to a critical temperature $T_{c}=22K$. For
La$_{2-x}$Sr$_{x}$CuO$_{4}$, a clear nodal gap is observed at
doping concentrations $x=0.01$ and $x=0.02$, but it closes at
$x=0.03$ which is less than the critical doping $x_{c}=0.05$ where
superconductivity starts to emerge as the ground state. In the
case of slightly electron-doped Nd$_{2-x}$Ce$_{x}$CuO$_{4}$, the
gap in the nodes is observed at doping $x=0.025$ and $x=0.04$, and
it does not exist at doping $x=0.08$ and $x=0.10$, well below the
critical doping concentration $x_{c}=0.12$. Such a doping
dependence of the excitation gap in the nodal direction is
observed in three different cuprate superconductors and hence
should be a universal phenomena. These observations are rather
striking because based on the $d$-wave symmetry of gap/pseudogap,
the energy gap should vanish along the nodal directions.

Although the nodal gap in the three cuprates has the same
dependence of the doping concentration, the inconsistency of
critical doping at which the nodal energy gap closes and critical
doping at which superconductivity emerges brings a difficulty to
understand the relationship between the nodal gap and high
temperature superconductivity. However, the difficulty is not that
severe as it seems to be, because all the above ARPES measurements
are performed at a finite temperature $T=15K$ rather than nearly
zero temperature. It is very probable that the true ground state
of all the lightly doped cuprates has the same electronic
structure and the above ARPES observed inconsistency is caused by
thermal fluctuations which are different from material to
material.

We speculate that this picture is what really happens in high
temperature superconductors. In the spirit of our results, at
doping concentration that is lower than the critical doping at
which superconductivity appears, the instability caused by strong
gauge fluctuations always generates a finite gap for the nodal
fermions. When superconductivity emerges, if its superfluid
density is less than some critical value, the nodal fermions also
have a finite gap. This finite gap is suppressed by a larger
doping concentration, i.e., a larger superfluid density. This is a
universal picture for the evolution of zero-temperature fermion
energy spectrum along the nodal direction in all cuprate
superconductors. However, the thermal fluctuations are different
in different materials and they destroy the spontaneously
generated nodal gap at different temperatures. This is
qualitatively in agreement with the ARPES data of Shen's group
\cite{kmShen}. To make quantitative comparison with experimental
data, detailed calculations considering chemical structure and
disorders are needed, which are beyond the scope of this paper.

In order to compare our results with experiments, the above
discussions on dynamical fermion gap generation should be extended
to finite temperatures. The problem has been investigated by
several authors with the results that above a critical temperature
the chiral symmetry is restored and the fermions remain gapless
\cite{Aitchison}. Before the holes are doped into the Cu-O plane,
the physical fermion flavor is $N_{f}=2$. According to the results
of Aitchison $et$ $al.$, $\frac{k_{B}T_{c}}{N_{f}}=0.002$, which
leads to $T_{c}=45K$. From Eq. (19) we know that the holons shift
the effective flavor to $N_{f}=3$. From the result of Aitchison
$et$ $al.$ we know that $\frac{k_{B}T_{c}}{N_{f}}$ is about
$10^{-4}$, which leads to $T_{c} \sim 4K$. The nodal gap
generation occurs only at temperatures below this critical value.
When the temperature is beyond this critical value, thermal
fluctuations destroy the nodal gap generation and hence the gap
along the nodal direction closes.

\subsection{Competition and coexistence of antiferromagnetism and superconductivity}

Understanding the competition of various ground states of cuprate
superconductors is one of the central problems in modern condensed
matter physics. The cuprate superconductors in its hall-filling
limit are believed to be Mott insulators with long-range
antiferromagnetic order. When the doping concentration increases,
long-range AF order becomes short-ranged and $d$-wave
superconductivity emerges as the ground state. It is interesting
to build a microscopic theory to describe the evolution from the
antiferromagnetism phase to the superconductivity phase with
doping.

The doping dependence of antiferromagnetism is very similar to
that of the nodal gap observed in ARPES measurements, indicating
that they might be governed by the same mechanism. This can be
easily verified by calculating the antiferromagnetic spin
correlation function. Once the nodal fermion acquires a finite
gap, the antiferromagnetic spin correlation is greatly enhanced.
Actually, it has been argued that the chiral symmetry breaking
corresponds to the formation of antiferromagnetic long-range order
\cite{Kim99, Tesanovic, LiuWF, Liu2003}. The gapless spin wave
excitation is interpreted as the massless Goldstone boson. The
antiferromagnetic spin correlation is defined as
\begin{equation}
\langle S^{+}S^{-}
\rangle=-\frac{1}{4}\int\frac{d^{3}k}{(2\pi)^{3}}Tr\left[G_{0}(k)G_{0}(k+p)\right],
\end{equation}
where $G_{0}(k)$ is the fermion propagator. If the fermions are
massless, then $G_{0}(k)=\frac{-i}{\gamma \cdot k}$ and we have
$\langle S^{+}S^{-} \rangle=-\frac{\left|p\right|}{16}$. At $p
\rightarrow 0$, $\langle S^{+}S^{-} \rangle_{0} \rightarrow 0$,
and the antiferromagnetic correlation is heavily lost. The
propagator for the massive fermion is
\begin{equation}
G(k)=\frac{-\left(\gamma \cdot
k+i\Sigma_{0}\right)}{k^{2}+\Sigma_{0}^{2}},
\end{equation}
where a constant mass $\Sigma_{0}$ is adopted which does not
affect the conclusion. This propagator leads to a correlation
function $\langle S^{+}S^{-}\rangle$ that behaves like
$-\Sigma_{0}/2\pi$ as $p \rightarrow 0$ and we have long-range
antiferromagnetic correlation when chiral symmetry breaking takes
place \cite{Liu2003}.

The gauge boson mass in the superconducting state is an
appropriate physical quantity to describe superconductivity. Thus
the relationship between antiferromagnetism and superconductivity
can be described by the relationship between spontaneous nodal gap
generation and the mass of gauge boson. We have shown that
spontaneous nodal gap generation can take place for doping
concentration less than $\delta_{c}$. This indicates that
antiferromagnetism only exists at half-filling and lightly doped
cuprates, in consistent with experiments. In most cuprate
superconductors, the superfluid density $\rho_{s}$ is large enough
to suppress antiferromagnetism once superconductivity emerges at
the critical doping concentration, which is generally larger than
$\delta_{c}$. Hence, antiferromagnetism can not coexist with
superconductivity in the bulk materials, at least in most cuprate
superconductors. However, if an external magnetic field is
introduced to the superconductors and reduces the superfluid
density $\rho_{s}$ down to below its critical value, then
antiferromagnetism intends to appear in the superconducting state.
The coexistence of antiferromagnetism and superconductivity is
thus possible.

Recently, intense investigation have been taken on the magnetic
field induced local antiferromagnetic order. The external magnetic
field perpendicular to the CuO$_{2}$ plane generates Abrikosov
vortices inside which the superfluid density is suppressed. Around
the vortex cores, if the superfluid density is reduced to below
the critical value, then spontaneous gap generation and hence
antiferromagnetic order can be formed in the vortex state
\cite{Lake, Mitrovic, Hoffman, Khaykovich}. Demler $et$ $al.$ have
attempted to explain these experiments by assuming a proximity to
the coexistence of spin-density wave and superconductivity
\cite{Demler}. Such a coexistence can also be addressed within
several other theories \cite{Zhang, Kishine, Franz, Zhu} including
Zhang's SO(5) theory \cite{Zhang} and the staggered flux approach
proposed by Lee and Wen \cite{Kishine}. In this paper we use
spin-charge separation and dynamical chiral symmetry breaking to
account for the competition and coexistence of antiferromagnetism
and superconductivity. Our approach emphasizes on the similarity
of this competition and coexistence with the transport behavior
and single particle spectrum properties. More quantitative
calculation will be carried out in the future in order to produce
the detailed experimental data.

\subsection{Low temperature heat transport behavior}

The low temperature transport behavior of superconductors is
controlled by the gap symmetry since it determines the type of
low-energy excitations. For conventional $s$-wave superconductors,
the low-energy density of states $N_{s}\left(\omega \right)$ damps
rapidly as temperature decreases, i.e., $N_{s}\left(\omega
\right)=0$ for $\left| \omega \right|<\Delta_{0}$ with
$\Delta_{0}$ the quantity of energy gap. However, the situation is
quite different for $d$-wave superconductors due to the presence
gap nodes. In the absence of impurities, the density of states is
$N_{d}\left(\omega \right) \sim \omega$. Due to the scattering of
impurities a finite density of quasiparticle states exist, i.e.,
$N_{d}\left(0 \right)$ is finite \cite{Durst}. Lee \cite{Lee93}
has investigated the quasiparticle transport of $d$-wave
superconductors and found that it is independent of impurity
intensity as a result of the competition between the growth of
quasiparticle density of states and the reduction of mean free
path. In contrast to $s$-wave superconductors where thermal
conductivity vanishes due to the absence of mobile fermions at
$T\rightarrow 0$, a linear term for thermal conductivity appears
in $d$-wave superconductors at $T\rightarrow 0$. Taillefer $et$
$al.$ for the first time observed such a linear term for thermal
conductivity in optimally doped cuprate YBa$_{2}$Cu$_{3}$O$_{6.9}$
\cite{Taillefer}. A similar linear term is also observed
\cite{Chiao} in optimally doped Bi$_{2}$Sr$_{2}$CaCu$_{2}$O$_{8}$.
Recently, extensive heat transport measurements have been
performed in La$_{2-x}$Sr$_{x}$CuO$_{4}$ and
YBa$_{2}$Cu$_{3}$O$_{y}$ in a wide range of doping concentration
from overdoping to very underdoping. A linear term for thermal
conductivity is observed through the whole superconducting region
\cite{Takeya, Sutherland}. Decreasing the doping concentration
drives the cuprate to the proximity to a critical point at which
spontaneous gap generation occurs. As a consequence, the thermal
conductivity should decrease down to zero at low doping. In the
spirit of our mechanism, the gapless nodal quasiparticles in the
superconducting state are stable against gauge fluctuations
because the gauge boson becomes massive and hence can not generate
a finite gap for the gapless nodal fermions. For doping
concentration below $\delta_{c}$, the nodal quasiparticles acquire
a finite gap which changes the node structure of $d$-wave gap
symmetry and suppresses the appearance of low-energy fermions.
Therefore, there is no linear term for the thermal conductivity at
very low doping region.

Recently the dependence of thermal transport on magnetic field has
been investigated. Naively, magnetic field should drive the
thermal conductivity to increase since it breaks the Cooper pairs
and hence quasiparticle density increases with magnetic field.
This is true for many superconductors including conventional
$s$-wave superconductors and overdoped cuprates. However, this is
not true for underdoped $d$-wave cuprate superconductors, in which
low temperature thermal conductivity decreases with magnetic
field. The crucial reason for this difference is that spin-charge
separation caused by strong electron correlation takes place in
underdoped cuprates but not in overdoped cuprates and conventional
BCS superconductors. When spin and charge degrees of freedom are
separated, the holon condensation is suppressed by strong magnetic
fields and consequently the mass of internal gauge boson decreases
down to zero with increasing magnetic field. On the contrary, the
spinon pairs are stable against the external magnetic fields.
However, the nodal quasiparticle is affected by the change of
internal gauge boson mass. When superconductivity is destroyed by
external magnetic fields at $H_{c2}$, the mass gap of internal
gauge boson vanishes. As a result, the gapless nodal fermions
acquire a finite gap. Below this gap the density of states of
quasiparticles is zero, preventing the appearance of low energy
fermion excitations. Thus we now see that, while magnetic filed
generates fermion quasiparticles in many ordinary superconductors,
the $d$-wave underdoped cuprate superconductors has a rather
peculiar property that the magnetic field reduces low-energy
fermions due to the spin-charge separation and spontaneous nodal
gap generation.

When superconductivity is completely suppressed by magnetic field,
heat is transported only by bosons including spin wave, holons and
phonons at low temperatures. These bosons can only contribute a
$T^{3}$ term to thermal conductivity at low temperature, which can
be nearly neglected. Therefore, according to our mechanism there
should be a thermal metal-to-insulator transition upon going from
the superconducting state to the field-induced normal state of
underdoped cuprates. This phenomenon has been observed in
underdoped La$_{2-x}$Sr$_{x}$CuO$_{4}$ recently in heat
measurements \cite{Hawthorn}. In contrast to the thermal insulator
property of the field-induced normal state, the holons can move
freely giving rise an metal-like charge transport behavior. Thus
we expect that the Wiedemann-Franz law, which gives a universal
relationship between the thermal conductivity $\kappa$ and the
electrical conductivity $\sigma$ as $\frac {\kappa}{\sigma
T}=\frac {\pi ^{2}}{3} \left( \frac {k_{B}}{e} \right )^{2}$ with
$k_{B}$ the Boltzmann's constant, should no longer hold in this
region. Experiments \cite{Hill} have provided certain evidence
supporting the breakdown of this law in the electron-doped cuprate
Pr$_{2-x}$Ce$_{x}$CuO$_{4-y}$.

\section{summary}

One remarkable property of high temperature superconductors is its
$d$-wave gap symmetry. Due to the presence of gap nodes, there is
an amount of low-energy quasiparticles which play a crucial role
in determining the low-temperature behavior of cuprate
superconductors. For example, the thermally excited nodal
quasiparticles can efficiently destroy the superfluid density
\cite{Lee97}. While at low temperatures, the nodal quasiparticles
contribute a finite thermal conductivity that is independent of
impurity concentrations. The $d$-wave gap in the single-particle
spectrum exists not only in the superconducting state but also in
the normal state of underdoped cuprates. The gapless nodal fermion
excitations couples to a strong gauge field and can acquire a
dynamically generated gap. When this happens, the gap nodes are
removed and all excitations are gapped. The spontaneous nodal gap
generation modifies the picture of low-temperature physics to a
new one, in which no free fermions can be found and a long-range
antiferromagnetic order is formed.

The spontaneous generation of nodal gap depends on the gauge
boson. If the gauge boson is massless, then the gap generation can
always take place. When superconductivity emerges, the gauge boson
becomes massive. There is a critical value for the gauge boson
mass, only below which could spontaneous gap generation for nodal
fermions take place. The high temperature superconductors have a
peculiarity that the gauge boson mass is proportional to the
doping concentration. Thus spontaneous nodal gap generation should
exist at low doping concentrations while vanishes as doping
increases. However, when superconductivity is suppressed by some
external means such as external magnetic field, spontaneous nodal
gap generation occurs and correspondingly low temperature nodal
quasiparticles disappear. Hence there is a thermal metal-insulator
transition from the superconducting ground state to the
field-induced normal ground state upon increasing magnetic field.
On the other hand, spontaneous nodal gap generation corresponds to
the formation of antiferromagnetic long-range order. The fate of
antiferromagnetism hence can be described by the relationship of
spontaneous nodal gap generation with superfluid density or doping
concentration.

\section{acknowledgement}

G.-Z. Liu would like to thank Prof. X.-H. Chen for helpful
discussions. G.-Z. Liu is supported by the Reinforcing Education
Project Towards 21th Century. G. Cheng is supported by National
Science Foundation in China No.10175058 and No.10231050.


\begin{thebibliography}{99}
\bibitem{note}
Some recent studies on QED$_{3}$ have arrived results against the
conclusion that the critival fermion flavor is about $3.3$.
Numerical simulations on lattice QED$_{3}$ performed by Hands $et$
$al.$ found no decisive signal for chiral symmetry breaking for $N
\geq 2$ \cite{Hands}. It was suggested by Gusynin and Reenders
\cite{Gusynin} that this result is owing to the introduction of
large infrared cutoff in lattice studies. Several years ago,
Appelquist $et$ $al.$ \cite{Appelquist99} argued that chiral
symmetry of QED$_{3}$ can not be broken if the flavor is larger
than $3/2$ based on a conjectured constraint on the infrared
structure of asymptotically free gauge theories. However, this
result is doubful because the reliability of this constraint is
not verified, as mentioned by Appelquist $et$ $al.$ in a very
recent review \cite{Appelquist04}. On the other hand, the original
result in Ref. (22) has been confirmed by extensive investigations
\cite{Nash, Maris} and now is widely accepted.
\end{thebibliography}
\end{document}